\documentclass[twocolumn,aps,floatfix,showpacs,10pt,prb]{revtex4-1}
\usepackage{amsmath}
\usepackage{graphicx}
\usepackage{epsfig}
\usepackage{graphics}
\usepackage{bm}
\usepackage{amssymb}
\usepackage{psfrag}
\begin{document}
\title{Orbital physics of polar Fermi molecules}
\author{Omjyoti Dutta$^1$, Tomasz Sowi\'nski$^{2}$, Maciej Lewenstein$^{1,3}$}
\affiliation{
\mbox{$^1$ ICFO ---Institut de Ciencies Fotoniques, Av. Carl Friedrich Gauss, num. 3, 08860 Castelldefels (Barcelona), Spain  }
\mbox{$^2$Institute of Physics of the Polish Academy of Sciences, Al. Lotnik\'ow 32/46, 02-668 Warsaw, Poland}
\mbox{$^3$ ICREA -- Instituci{\'o} Catalana de Recerca i Estudis Avan\c{c}ats, Lluis Companys 23, E-08010 Barcelona, Spain} }
\date{\today}
\begin{abstract}
We study a system of polar dipolar fermions in a two-dimensional optical lattice and show that multi-band Fermi-Hubbard model
is necessary to discuss such system. By taking into account both on-site, and long-range interactions between different bands,
as well as occupation-dependent inter- and intra-band tunneling, we predict appearance of novel phases in the strongly-interacting limit.
\end{abstract}
\maketitle
\section{Introduction}
Creation of ultracold hetero-nuclear molecules opens the path towards
experimental realization of strongly-interacting dipolar many-body
systems. Depending on the constituent atoms, in moderate electric field
these molecules can have large dipole moment of $1$ Debye in their vibrational ground states \cite{Jin, LiCs, Nag, Zwr}.
In particular, fermionic molecules in presence of an optical
lattice can be used to simulate various quantum phases, such as quantum
magnetism and phases of $t-J$ like models \cite{Gorshov, Rey}, various
charge density wave orders \cite{Free, Bruun}, bond-order solids
\cite{Bhong} etc. One should also stress that in the strongly correlated regime,
both in bosonic and fermionic systems the standard descriptions of single-band Hubbard model ceases to be valid.
The effect of non-standard terms become important leading to novel phases like pair-superfluidity etc \cite{Dirk1, Dirk2, Mer, Hof, Dutta, Tom}.

While most of the works have dealing with higher bands concentrated
on bosonic systems, in this paper, we study dipolar fermions
confined in 2D optical lattice $V_{\mathrm{latt}}=
V_0\left[\sin^2(\pi x/a) + \sin^2(\pi y/a)\right] +
\frac{m\Omega^2}{2}z^2$, where $V_0$ is the lattice depth, $a$ is
the lattice constant, $m$ is the mass of the molecule, and $\Omega$
is the frequency of harmonic potential in $z$ direction. The dipoles
are polarized along the direction of harmonic trapping. Usually, at
low temperature and for low tunneling, the phase diagram consists of
different crystal states whose structure depends on the filling $n$
\cite{Free}. In this paper, we derive a Fermi-Hubbard model for
dipolar fermions including the effects of higher bands. We show
that, even for moderate dipolar strength, it is necessary to take
into account the excitations along the $z$ direction.
Simultaneously, in this regime, the interaction induced hopping
along the lattice give also important contributions. This changes
the phases expected for a spinless Hubbard model including only a
single band. Near $n\gtrsim 1/4$, we find a spontaneous appearance
of non-Fermi liquid behaviour in the form of smectic metallic phase.
Near $n\gtrsim1/2$, we find that the system can be mapped to an
extended pseudo-spin $1/2$ Hubbard model with different emergent
lattice configuration. We find a regime where chiral $p$-wave
superconductivity emerges through Kohn-Luttinger (KL) mechanism with
transition temperature $T_c$ of the order of tunneling. This gives
rise to an exotic supersolid, with the diagonal long-range order
provided by the checkerboard pattern of the lower orbital fermions,
while the superfluidity originating from the fermions in the higher
band.

The paper is organized as follows : In section II we have introduced a multi-orbital model
to describe dipolar fermions in optical lattices. We then discuss quantitatively the contributions of different
parameters present in the model. In section III we have described the energy contribution of different crystal structures in the limit of vanishing tunneling. We also compare the corresponding energies of such crystal states without taking into account the higher bands and show that it is necessary to take into account the higher band contributions for experimentally realizable parameters. In section IV, we have investigated the ground state properties for filling greater than $1/4$. We find that due to the higher band occupation dependent tunneling contributions, within certain parameter regime, there is a spontaneous formation of smectic-metal phase, along with stripe-like phases. In section V we describe the ground state structures for $n\gtrsim 1/2$. We find that the higher-band tunneling can give rise to
sub-lattices which further can give rise to $p$-wave superfluidity. In section VI we present our conclusions followed by acknowledgements in section VII.

\section{Model}

The Hamiltonian for the dipolar fermions in the second quantized
form reads $H=\int\mathrm{d}^3\mathbf{r}
\Psi^\dagger(\mathbf{r})H_0\Psi(\mathbf{r}) +
\frac{1}{2}\int\mathrm{d}^3\mathbf{r}\,\mathrm{d}^3\mathbf{r}'
\Psi^\dagger(\boldsymbol{r})\Psi^\dagger(\boldsymbol{r}'){\cal
V}(\mathbf{r-r'}) \Psi(\mathbf{r}')\Psi(\mathbf{r}),$ where
$\Psi(\mathbf{r})$ is a spinless fermion field operator. In the
units of recoil energy $E_R = \pi^2\hbar^2/(2 m a^2)$, the single
particle Hamiltonian becomes $H_0=- \nabla^2 + V_{\rm
latt}(\mathbf{r})/E_R$ and the long-rage interaction potential
${\cal V}(\mathbf{r})= D \left(1/r^3-3z^2/r^5\right)$, where $D =
2\pi m d^2 /(\hbar^2 a )$ is a dimensionless dipolar strength,
related to the electric dipolar moment $d$. For {\rm KRb} molecules
with a dipole moment of $0.5$ Debye confined in the optical lattice
with $a=345{\rm nm}$ \cite{Ties} one gets $D=8.6$ whereas, for
similar lattice parameters, {\rm LiCs} molecules can have a dipole
moment of $\sim 5$ debye with $D\sim 100$. We decompose the field
operator in the basis of Wannier functions in the $x, y$ directions
and of harmonic oscillator eigenstates in $z$ direction. For
convenience we introduce orbital index $\sigma=\{pml\}$ denoting
$p$, $m$ and $l$ excitations in $x$, $y$, and $z$ direction
respectively. In this basis the field operator $\Psi(\mathbf{r} ) =
\sum_{\boldsymbol{i},\sigma} \hat{a}_{\boldsymbol{i}\sigma} {\cal
W}_{\boldsymbol{i}\sigma}(\boldsymbol{r})$, where ${\cal
W}_{\boldsymbol{i}\sigma}(\boldsymbol{r})$ is the single-particle
wave-function in orbital $\sigma$ localized on site $\boldsymbol{i}=
i_x \boldsymbol{e}_x+i_y\boldsymbol{e}_y$ ($\boldsymbol{e}_x$ and
$\boldsymbol{e}_y$ are unit vectors in the proper directions).
Fermionic operator $\hat{a}_{\boldsymbol{i}\sigma}$ annihilates
particle in this state. The Hamiltonian can be rewritten in the
following Hubbard-like form $H=\sum_\sigma {\cal
H}_\sigma^{(1)}+\sum_{\sigma\sigma'}{\cal H}_{\sigma\sigma'}^{(2)}$
where
\begin{subequations} \label{HHubbard}
\begin{align}
{\cal H}_\sigma^{(1)} &= E_\sigma\sum_{\boldsymbol{i}} \hat{n}_{\boldsymbol{i}\sigma} + J_\sigma \sum_{\{\boldsymbol{ij}\}} \hat{a}_{\boldsymbol{i}\sigma}^\dagger \hat{a}_{\boldsymbol{j}\sigma} \\
{\cal H}_{\sigma\sigma'}^{(2)} &=  U_{\sigma\sigma'} \sum_{\boldsymbol{i}} \hat{n}_{\boldsymbol{i}\sigma}\hat{n}_{\boldsymbol{i}\sigma'} +\sum_{\boldsymbol{i}\neq\boldsymbol{j}} V_{\sigma\sigma'}(\boldsymbol{i}-\boldsymbol{j})\hat{n}_{\boldsymbol{i}\sigma}\hat{n}_{\boldsymbol{j}\sigma'} \nonumber \\
&+ \sum_{\{\boldsymbol{ij}\}}\sum_{\sigma''} T_{\sigma\sigma'}^{\sigma''}(\boldsymbol{i-j})\hat{a}_{\boldsymbol{i}\sigma}^\dagger \hat{n}_{\boldsymbol{i}{\sigma''}}\hat{a}_{\boldsymbol{j}\sigma'}. \label{HHubbardB}
\end{align}
\end{subequations}
Parameters $E_\sigma$ and $J_\sigma$ comes from the single particle
Hamiltonian and denote single-particle energy and nearest-neighbour
tunneling in orbital $\sigma$ respectively. The inter-particle
interaction has three contributions to the Hamiltonian
\eqref{HHubbardB} : {\it (i)} the on-site interaction energy of
fermions occupying different orbitals $\sigma$ and $\sigma'$ of the
same site $U_{\sigma\sigma'}$, {\it (ii)} the long-range interaction
energy of fermions occupying orbitals $\sigma$ and $\sigma'$ of
different sites $V_{\sigma\sigma'}(\boldsymbol{i-j})$, {\it (iii)}
and the tunneling from orbital $\sigma'$ at site $\boldsymbol{j}$ to
the orbital $\sigma$ at site $\boldsymbol{i}$ induced by presence of
an additional fermion at site $\boldsymbol{i}$ in orbital $\sigma''$
denoted by $T^{\sigma''}_{\sigma\sigma'}(\boldsymbol{i-j})$.

The Hamiltonian \eqref{HHubbard} is very general. To get a physical
understanding of its properties, we start by examining the
properties of density-density interactions. We calculate the
interactions between few lowest bands: $s=\{000\}$, $p_x=\{100\}$,
$p_y=\{010\}$, $p_z=\{001\}$, $p_{xz}=\{101\}$, and
$p_{yz}=\{011\}$. We find that the on-site interactions
$U_{s,p_x}=U_{s,p_y}$ is always repulsive. It means that putting two
fermions in $s$ and $p_x$ or $p_y$ band simultaneously is
energetically unfavorable. Remarkably we find that $U_{s,p_z}$ is
always negative. This surprising attraction stems from the presence
of the fermionic exchange term and the shape of the dipolar
interactions (see Appendix A.). Moreover, as it is seen
from Fig. \ref{interaction}a, this interaction can not be neglected
even for stronger confinements in $z$ directions. For higher
orbitals we find that $U_{s,p_z}\ll U_{s,p_{xz}} = U_{s,p_{yz}} <0
$. In addition for long-range interactions we find that
$V_{s,s}(\boldsymbol{i})>V_{s,p_z}(\boldsymbol{i})>V_{p_z,p_z}(\boldsymbol{i})>0$.
From this analysis we conclude that for polar molecules there always
exists some critical dipolar strength for which single-band
approximation breaks down since two particles can occupy the same
site. This critical behavior is controlled by the on-site energy
cost $\Delta = E_z + U_{s,p_z}$ where the energy gap between $s$ and
$p_z$ orbital is given by $E_z=\hbar\Omega$. We find that
$U_{s,p_z}\sim -DE_R\sim-\hbar\Omega$ for $D\sim8$. Thus it is
important to take into account atleast $s$ and $p_z$ orbital to
describe the dipolar fermions. In this paper we consider situations
when $\Delta$ is positive due to which three or more fermions in
same site is unfavourable.

\begin{figure}
\includegraphics{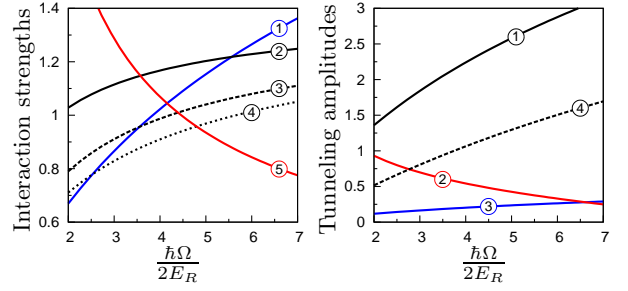}
\caption{Parameters of the Hamitlonian for $V_0=8E_R$ as functions of the lattice confinement. (a) On-site interaction $-\frac{U_{s,p_z}}{DE_R}$ (blue solid line -1-), and nearest-neighbor long-range interactions $\frac{V_{s,s}\boldsymbol{e}_x)}{DJ_s}$ (solid black line -2-), $\frac{V_{s,p_z}(\boldsymbol{e}_x)}{DJ_s}$ (dashed black line -3-), and $\frac{V_{p_z,p_z}(\boldsymbol{e}_x)}{DJ_s}$ (dotted black line -4-). The red line -5- shows the ratio $V_{p_z,p_z}(2\boldsymbol{e}_x)/T_{\mathtt{eff}}^{\parallel}$. (b) Magnitudes of the induced tunneling terms
$\frac{T_{p_z,p_{xz}}^{s}(\boldsymbol{e}_x)}{DJ_s}$ (black solid line -1-), $\frac{T_{s,p_x}^{s}(\boldsymbol{e}_x)}{DJ_s}$ (red solid line -2-), and $\frac{T_{p_z,p_z}^{s}(\boldsymbol{e}_x)}{DJ_s}$ (blue solid line -3-). The dashed black line -4- denotes the ratio $T_{\mathtt{eff}}^{\parallel}/J_s$ for $D=10$.} \label{interaction}
\end{figure}

Next we discuss the role of the interaction induced tunnelings in
Hamiltonian \eqref{HHubbard}. Counter intuitively the most important
contribution does not come from the induced tunneling in $p_z$ band
($T_{p_z,p_z}^s(\boldsymbol{e}_x)$), but from the inter-band
tunneling which changes $p_z$ orbital to the $p_{xz}$ and $p_{yz}$
ones ($T_{p_z,p_{xz}}^s(\boldsymbol{e}_x)$). Note, that this
inter-band tunneling is absent for usual single-particle tunneling
due to the properties of single particle Hamiltonian \cite{Kohn}.
From properties of $p$ orbital states it follows that
$T_{p_z,p_{xz}}^s(-\boldsymbol{e}_x)=-T_{p_z,p_{xz}}^s(\boldsymbol{e}_x)$.
The relation of this term to other interaction-induced tunnelings is
shown in Fig. \ref{interaction}b. From the above analysis we
introduce simplified, but realistic model of polar Fermi molecules
confined in 2D optical lattice by taking into account effects of
interactions between orbitals $\sigma\in \{s,p_z,p_{xz},p_{yz}\}$.

\section{Comparison of energies between different ground state candidates}

To obtain an idea about the ground state structures, in this section we compare the energies of different possible ground-state crystal configurations for specific filling factors with and without the contributions from the $p_z$-orbitals.
%---------------------------
\begin{figure}
\includegraphics[scale=0.4]{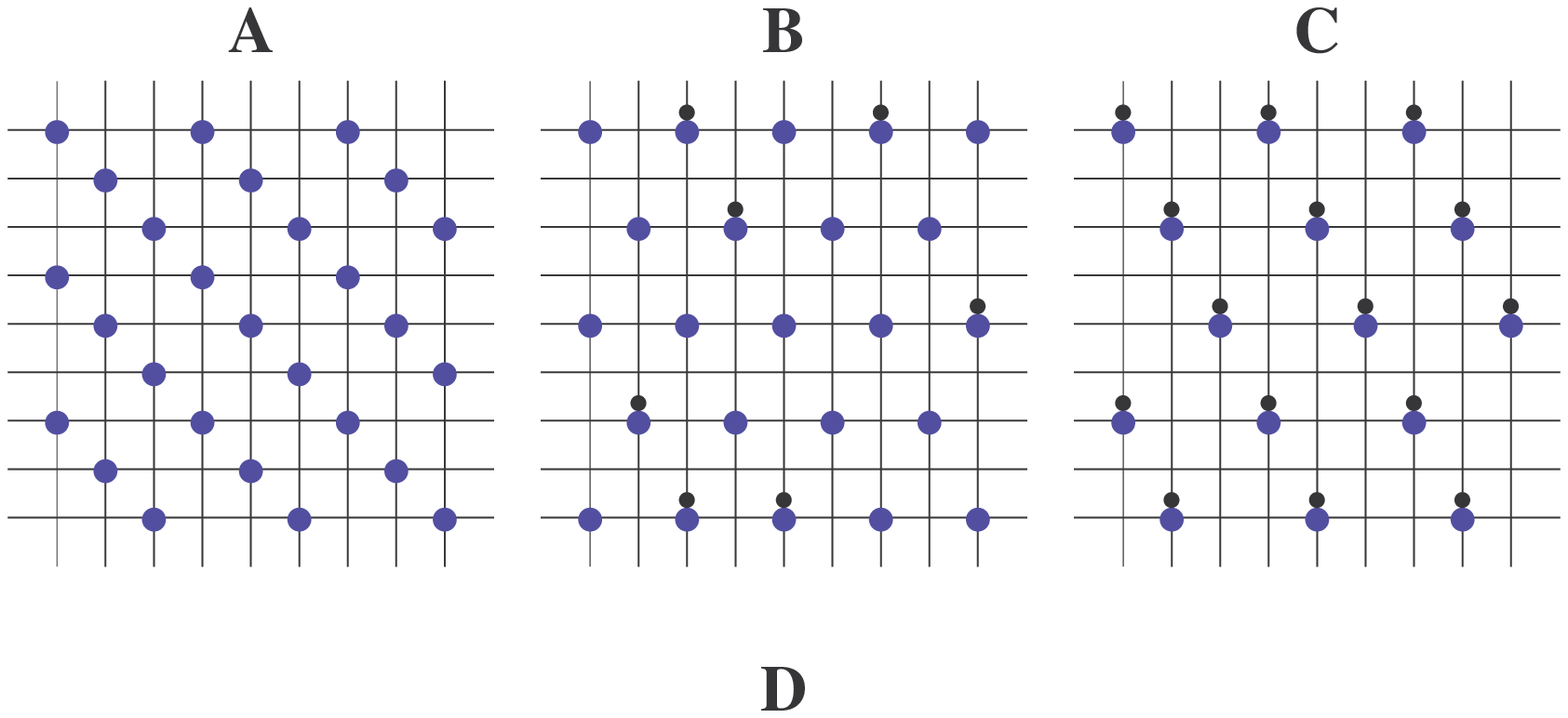}
\psfrag{xlabel}{$\frac{\hbar\Omega}{2E_R}$}
\includegraphics[scale=0.5]{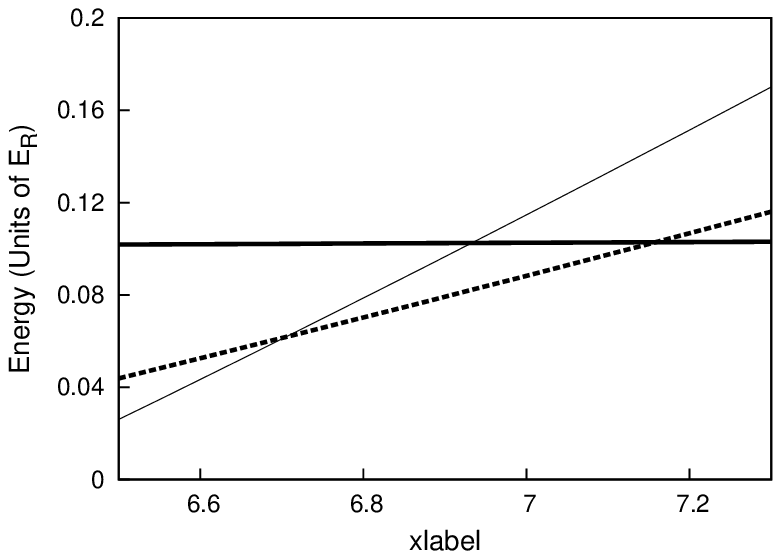}
\caption{\label{lattice1} Pictorial diagram of the different checker board lattices for $n=1/3$. The blue spheres denote
$s$-orbital fermions and the smaller black spheres denote $p_z$ orbital fermions. (A) Ground state crystal phase of Hamiltonian \eqref{HamI}. (B) $1/4$ checkerboard lattice of $s$-band fermions and extra $p_z$ fermions with density $1/12$. (C) Density-wave structure of the effective bosons with filling $n^b=1/6$ corresponding to the ground state structure of
the Hamiltonian \eqref{HamIII}. (D) The energies $E_{2A}$ (thick-solid line), $E_{2B}$ (dashed line), and $E_{2C}$ (thin-solid line) as functions of the trap frequency $\hbar\Omega/2E_R$ for dipolar strength $D=10$.}
\end{figure}
%----------------------------
For the clarity of discussion, we first neglect the tunneling terms as justified in the strongly coupled regime. Without the higher orbital effects at most one fermion can occupy a given site. The corresponding Hamiltonian reads:
\begin{equation}\label{HamI}
{\cal H}_{I}=\sum_{\boldsymbol{i}\neq\boldsymbol{j}} V_{ss}(\boldsymbol{i}-\boldsymbol{j})\hat{n}_{\boldsymbol{i}s}\hat{n}_{\boldsymbol{j}s}
\end{equation}
with the dipolar interaction in the $s$ band $V_{ss}(\boldsymbol{i}-\boldsymbol{j})=V_{ss}/|\boldsymbol{i}-\boldsymbol{j}|^3$.

By taking the orbital effects into account, the corresponding Hamiltonian is defined in Eq. (1a) and (1b),
\begin{eqnarray}\label{HamII}
{\cal H}_{II} &=&  E_\sigma\sum_{\boldsymbol{i}} \hat{n}_{\boldsymbol{i}\sigma}+\sum_\sigma U_{\sigma\sigma'} \sum_{\boldsymbol{i}} \hat{n}_{\boldsymbol{i}\sigma}\hat{n}_{\boldsymbol{i}\sigma'} \nonumber\\
&+& \sum_{\sigma, \sigma'} \sum_{\boldsymbol{i}\neq\boldsymbol{j}} V_{\sigma\sigma'}(\boldsymbol{i}-\boldsymbol{j})\hat{n}_{\boldsymbol{i}\sigma}\hat{n}_{\boldsymbol{j}\sigma},
\end{eqnarray}
where $\sigma$ denotes the $s$- and $p_z$ orbital fermions.

Now we consider the situation, when each occupied site contains two fermions. In this case we can define a corresponding hardcore bosonic operator at site $\boldsymbol{i}$ as
$\hat{b}^{\dagger}_{\boldsymbol{i}}=s^{\dagger}_{\boldsymbol{i}}p^{\dagger}_{z\boldsymbol{i}}$ and $\hat{b}_{\boldsymbol{i}}=p_{z\boldsymbol{i}}s_{\boldsymbol{i}}$ and the
bosonic number operator $\hat{n}^b_{\boldsymbol{i}}=\hat{b}^{\dagger}_{\boldsymbol{i}}\hat{b}_{\boldsymbol{i}}$. From this we can see that $n^b=n/2$ as the number of fermions is twice the number of bosons. Subsequently, we can write an effective bosonic Hamiltonian as:
\begin{equation}\label{HamIII}
{\cal H}_{III} =  \Delta \sum_{\boldsymbol{i}} \hat{n}^b_i + \sum_{\boldsymbol{i}\neq\boldsymbol{j}} \sum_{\sigma,\sigma'} V_{\sigma\sigma'}(\boldsymbol{i}-\boldsymbol{j})\hat{n}^b_{\boldsymbol{i}}\hat{n}^b_{\boldsymbol{j}},
\end{equation}
where $\sigma,\sigma'=s,p_z$. Here $\Delta = E_z + U_{s,p_z}$ is the energy cost of having a composite boson. Eq. \eqref{HamIII} is similar to the bosonic dipolar system with modified dipolar interaction and can simulate the crystal phases of dipolar bosons \cite{cap}.

For concreteness we first specifically choose $n=1/3$. At filling $n=1/3$ the ground state of the single band Hamiltonian \eqref{HamI} forms a crystal structure in accordance with \cite{Free} and it is shown in Fig.~\ref{lattice1}A. Its energy is $E_{2A}$. In the current paper, we analyze other structures as a ground states corresponding to the full Hamiltonian \eqref{HamII}. Two such structures are presented in Fig.~\ref{lattice1}B and \ref{lattice1}C with corresponding energies $E_{2B}$ and $E_{2C}$. In the $2B$ structure the $s$-band fermions form a $1/4$ crystal structure and remaining $1/12$ $p$-orbital fermions occupy already occupied sites. The third possible ground state candidate Fig.~\ref{lattice1}C comes from the effective bosonic Hamiltonian \eqref{HamIII} at filling $n_b=1/6$. We compare energies of these three structures by plotting them as functions of the harmonic trapping frequency for a dipolar strength $D=10$ (Fig.~\ref{lattice1}D).

\begin{figure}
\includegraphics[scale=0.5]{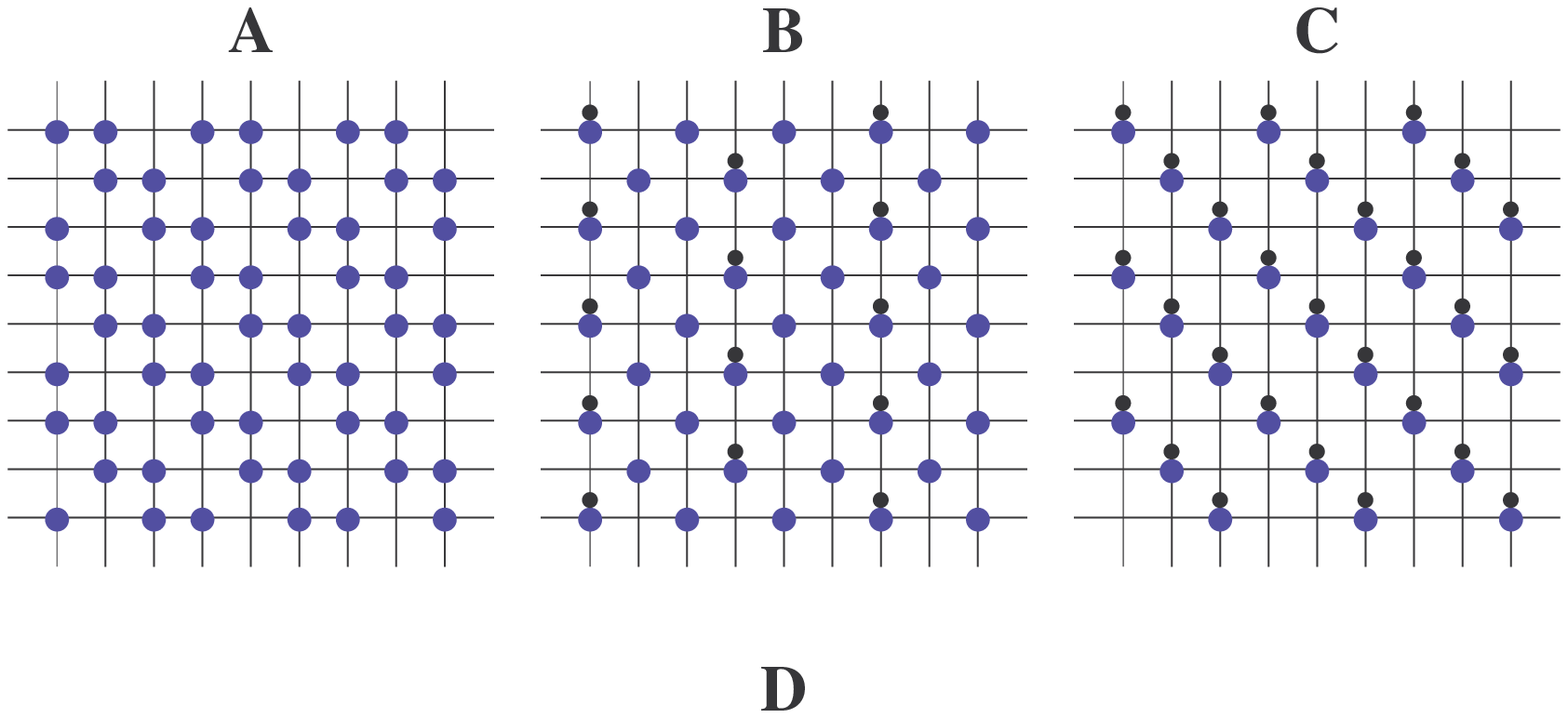}
\psfrag{xlabel}{$\frac{\hbar\Omega}{2E_R}$}
\includegraphics[scale=0.5]{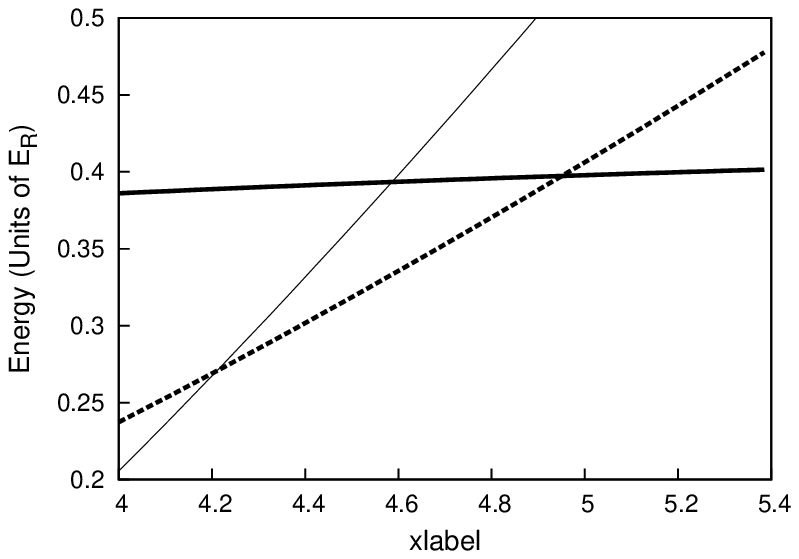}
\caption{\label{lattice2} Pictorial diagram of the different checkerboard lattices for $n=2/3$. The blue spheres denote
$s$-orbital fermions and the smaller black spheres denote $p_z$ orbital fermions. (A) Ground state crystal phase of Hamiltonian \eqref{HamI}. (B) $1/2$ checkerboard lattice of $s$-band fermions and extra $p_z$ fermions with density $1/6$. (C) Density-wave structure of the composite bosons with filling $n^b=1/3$ corresponding to the ground state structure of the Hamiltonian \eqref{HamIII}. (D) The energies $E_{3A}$ (thick-solid line), $E_{3B}$ (dashed line), and $E_{3C}$ (thin-solid line) as functions of the trap frequency $\hbar\Omega/2E_R$ for dipolar strength $D=8$.}
\end{figure}

We find that the energy of the structure $2A$ is almost insensitive to the trapping frequency $\Omega$. Moreover, the structure is the lowest energy state (the true ground state of the system) only for large enough $\Omega$ ($\hbar\Omega \gtrsim 14.5 E_R$ for studied case). For lower trap frequencies we find that structure $2B$ ($13.2 E_R\lesssim\hbar\Omega \lesssim 14.5 E_R$) or $2C$ ($\hbar\Omega\lesssim 13.2E_R$)  becomes a ground states of the system. We also note that in the structure $2C$, within the bosonic subspace, tunneling can arise in second order processes and it is much lower than the binding energy of the bosons. We have also checked that, for filling factors between $n=1/4$ and $n=1/3$, the energy of the configuration $2B$ is lower than the energy of the phase-separated structures of single-band Hamiltonian. Similarly we can infer also the ground state structures at filling $n=2/3$ as the situation is very simmilar to the filling $n=1/3$. The ground state of the single band Hamiltonian \eqref{HamI} shown in Fig.~\ref{lattice2}A with corresponding energy $E_{3A}$ is a true ground state of the system only for large enough $\Omega$. For lower confinement frequencies the
ground state is {\it (i)} a $1/2$ checkerboard $s$-band crystal with $p$-band fermions (with density $1/6$) moving on the occupied sites
(energy $E_{3B}$ and Fig.~\ref{lattice2}B) or {\it (ii)} $n_b=1/3$ stripe structure of composite bosons (energy $E_{3C}$ and Fig.~\ref{lattice2}C) \cite{cap}.
%---------------------------
%----------------------------

Similar results are also obtained for other filling fractions, namely $n=1/4,1/2,3/4$. For these filling fractions we also find that below a certain critical trapping strength $\Omega$, for critical $D$, it is important to take into account the excited trap states.

\section{Ground state structures near $n\gtrsim 1/4$}

In this section we will look into the properties of the ground states near $n=1/4$ filling.
We show that the presence of higher orbitals not only changes the ground-state crystal structures, it also fundamentally changes the properties of such states. Specifically we show that new forms of matter, like smectic metal
phase, can spontaneously form due to the effect of higher orbitals.

First, here we consider the case when $\Delta>0$, therefore for low
filling all fermions occupy only the $s$ orbital states. For filling
$n=1/4$ and large enough $D$ ($\gtrsim 3$) there is non vanishing
single-particle excitation gap and the system is in the $s$-band
insulator state (denoted by blue spheres in Fig.\ref{lattice}a)
\cite{Free}. Situation change dramatically for higher fillings. It
can be simply understood with energy arguments. The energy cost of
putting additional particle in the vacant site is given by
$E_{\mathtt{vac}}=
V_{s,s}(\boldsymbol{e}_x)+2V_{s,s}(\boldsymbol{e}_x+\boldsymbol{e}_y)+\ldots$.
In contrast the cost of putting additional particle to the $p_z$
orbital of an occupied site
$E_{\mathtt{occ}}=\Delta+2V_{s,p_z}(2\boldsymbol{e}_x)+\ldots$. For
$D$ larger than some critical strength one finds that
$E_{\mathtt{occ}}<E_{\mathtt{vac}}$. As an example, such conditions
are fulfilled for $V_0=8E_R$, $D=10$, and $\hbar\Omega \leq 14E_R$.
Consequently, additional particles start to fill $p_z$ band of
previously occupied sites. In this scenario energy conserving
dynamics of the system comes from the second-order processes
involving tunneling to the next occupied site (along $x$ direction
in Fig.~\ref{lattice}a). To the leading order, this effective
tunneling is given by
\begin{equation} \label{TunnelPar}
T_{\mathtt{eff}}^{\parallel} \approx T_{s,p_z}^s(\boldsymbol{e}_x)^2 /(|U_{s,p_z}|+E_x).
\end{equation}
Thus, the $p_z$ fermions will only move along one direction chosen
by the insulator checkerboard geometry in $s$-band, in our case
along $\boldsymbol{e}_x$. The resulting system can be thought as
stacks of one-dimensional chains or stripes placed along
$\boldsymbol{e}_y$ without inter-chain tunnelings.
The effective Hamiltonian governing the $p_z$ fermions can be
written as  $H_{\mathtt{1D}}=T_{\mathtt{eff}}^{\parallel} \sum_l
\sum_{\langle ij\rangle} \hat{c}^\dagger_{l,i}\hat{c}_{l,j}+H_{\mathtt{intra}}+H_{\mathtt{inter}}$
with intra-chain Hamiltonian $H_{\mathtt{intra}}=\sum_l\sum_{i,j} V_{\mathtt{intra}}(i,j)
\hat{c}^\dagger_{l,i}\hat{c}_{l,i}\hat{c}^\dagger_{l,j}\hat{c}_{l,j}$ and
inter-chain Hamiltonian $H_{\mathtt{inter}}= \sum_{l,l'} \sum_{i,i'} V_{l,l'}(i,i')\hat{c}^\dagger_{l,i}\hat{c}_{l,i}\hat{c}^\dagger_{l',i'}\hat{c}_{l',i'},$
where $\hat{c}^\dagger_{l,i_x}$ and $\hat {c}_{l,i_x}$ are creation and annihilation
operators of $p_z$ fermions on $s$-fermion occupied site $i$ on chain $l$. The intra-chain and inter-chain
interactions are given by $V_{\mathtt{intra}}(i,j)=V_{p_z,p_z}([i-j]\boldsymbol{e}_x)$ and
$V_{l,l'}(i,i')=V_{p_z,p_z}([i-i']\boldsymbol{e}_x+[l-l']\boldsymbol{e}_y)$ respectively.

The ground state structure of this coupled-chains system is
investigated by introducing the bosonized fields $\phi_{l,R/L}$ related to the Fermi operator
$\hat{c}_{l,i}$ rewritten in the continuum limit
as $\hat{c}_{l,i} \rightarrow \Psi_{l,L}(x) + \Psi_{l,R}(x)$ \cite{Emery, Vish}.
Near the left and right Fermi momenta $\pm\tilde{k}_1$, we can write $\Psi_{l,R/L}(x)=F_{R/L}\exp[{\pm}i\tilde{k}_1x-i\phi_{l, R/L}(x)]/\sqrt{2\pi\epsilon}$, where $\epsilon$ is a cutoff length and $F_{R/L}$ are Klein factors.
The Fermi momentum is given by the density of $p_z$ fermions which in terms of total density $n$ reads,
$\tilde{k}_1\approx(4n-1)\pi$. By writing the bosonized phase field $\theta_l(x)=(\phi_{l,L}(x)-\phi_{l,R}(x))/2\sqrt{\pi}$
in terms of its Fourier transform  $\theta_{q_y}(x)$ along the $\boldsymbol{e}_y$ the Lagrangian for the system reads
\begin{equation} \label{Lag1}
{\cal L}= \int^{\pi}_{-\pi} \frac{dq_y}{2\pi} \frac{K(q_y)}{2}
\left[\frac{1}{v(q_y)} \left(\frac{\partial\theta_{q_y}}{\partial t}\right)^2-v(q_y)\left(\frac{\partial\theta_{q_y}}{\partial x}\right)^2\right].
\end{equation}
The interaction parameter $K(q_y)$ and sound velocity $v(q_y)$
are determined by the details of the dipolar interactions (see appendix C).

There is an additional inter-chain $p_z$ fermion CDW coupling ${\cal L}_{\rm CDW}\propto \cos (\tilde{k}_1)\sum_{l}
\cos\sqrt{\pi}(\theta_l-\theta_{l+1})$. Consequently, the scaling dimension of
the CDW operator is given by $\eta_{l}=2\int^{\pi}_{-\pi}
\frac{1-\cos{lq_y}}{K(q_y)}\frac{dq_y}{2\pi}$. When $\eta_{l}>2$ the CDW operator is irrelevant.
Then the stable phase has properties similar to 1D Luttinger liquid with low-energy bosonic
collective excitations. This state preserves the smectic symmetry
$\theta_l\rightarrow\theta_l+\alpha_l$, with $\alpha_l$ constant on each chain.
This phase is known as {\it smectic-metal} phase \cite{Emery} as there metallic behavior along the chain with insulating
density wave order along transverse direction. This phase is a peculiar example of spontaneous emergence
of non-Fermi liquid behaviour in two-dimensional Fermi systems. In contrast, when $\eta_{l}<2$ then $p_z$ fermions becomes unstable towards formation of stripe crystals. In Fig.~\ref{lattice}b we plot $\eta_{1}$
and $\eta_{2}$ as functions of total density $n$ for $D=10$ and $\hbar\Omega=14E_R$.
It is clear that for $1/4<n<n_c$ there is a {\it smectic-metallic} phase while for $n>n_c$, the system goes to a stripe-crystal phase.

\section{Ground state structures near $n\gtrsim 1/2$}

\begin{figure}
\includegraphics[scale=0.7]{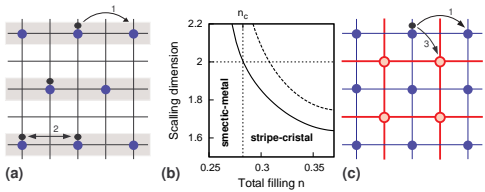}
\caption{\label{lattice} Pictorial diagram of the
different checker board lattices. The filled-blue and open-red spheres
denotes $s$-orbital fermions and the smaller black sphere denotes
$p_z$ orbital fermions. (a) Checkerboard lattice at $n=1/4$ filling.
The $p_z$ fermions will move with effective tunneling
$T_{\mathtt{eff}}^{\parallel}$ (arrow 1) only along the shaded
region making a stack of 1D chains. Interaction between neighboring
$p$-band fermions is equal to $V_{p_z,p_z}(2\boldsymbol{e}_x)$
(arrow 2). (b) Scaling dimensions $\eta_{1}$ (solid-line) and $\eta_{2}$ (dashed-line) as functions of
the total density $n$ for $D=10$ and $\hbar\Omega=14E_R$. (c) Checkerboard lattice at $n=1/2$. The blue and
thick-red lines constitute two different sub-lattices. They are not
coupled via tunneling processes since the tunneling
$T_{\mathtt{eff}}^{\perp}$ (arrow 3) is much smaller than
$T_{\mathtt{eff}}^{\parallel}$. }
\end{figure}

\begin{figure}
\includegraphics[scale=0.9]{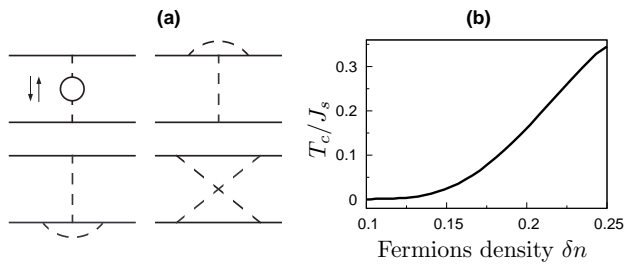}
\caption{ (a) Diagrammatic representation of the second order contributions in (\ref{effV}). The dashed lines denote interaction and the solid lines denote fermion propagator. (b) $p$-wave superfluid transition temperature $T_c$ as a function of density $n=1/2+\delta n$.  \label{Diagrams}}
\end{figure}

In this section let us discuss the case of filling $n=1/2$ where for low
dipolar strength $D$, due to the same reasons as before fermions
will occupy only the $s$-band and the ground state of the system is
the checkerboard insulator (see Supplimentary sec. D.)
as denoted by filled-blue and open-red spheres in Fig. \ref{lattice}b.
%-------------------------------
%-------------------------------------
To look for properties of the system with additional particles, we
define deviation from half-filling $\delta n = n -1/2$  and
we introduce corresponding chemical potential $\mu(\delta n)$. From
energy arguments we find that two scenario can happen. The
additional fermion {\it (i)} occupies a vacant site with energy cost
$E_{\mathtt{vac}} =
4V_{s,s}(\boldsymbol{e}_x)+8V_{s,s}(2\boldsymbol{e}_x+\boldsymbol{e}_y)
+ \ldots$ or {\it (ii)} it goes to the $p_z$ orbital of an occupied
site with energy cost $E_{\mathtt{occ}}=\Delta+4
V_{s,p_z}(\boldsymbol{e}_x+\boldsymbol{e}_y)+V_{s,p_z}(2\boldsymbol{e}_x)
+ \dots$. Consequently, in the second scenario (when
$E_{\mathtt{occ}} \le E_{\mathtt{vac}}$), all extra fermions will
occupy the $p_z$ orbitals of the already occupied sites. As an
example, such conditions are fulfilled for $V_0=8E_R$, $D=8$, and
$\hbar\Omega \leq 10E_R$. In such a case $\delta n$ corresponds to
the filling of $p_z$ band fermions. The parallel tunneling of the
$p_z$ fermions between the occupied sites will again arise from the
second order processes \eqref{TunnelPar}. Moreover, tunneling to the
diagonally occupied site $T_{\mathtt{eff}}^{\perp} \approx -\left [
J_s-T_{p_z,p_z}^s(\boldsymbol{e}_x) \right]^2/|U_{s,p_z}|$ for $D
\sim 8$ it is $400$ times smaller than
$T_{\mathtt{eff}}^{\parallel}$. Consequently fermions in the $p_z$
orbitals can move in independent square sub-lattices (either the
thick-red or blue sub-lattice shown in the Fig.~\ref{lattice}b).
Note, that fermions can not tunnel between different sub-lattices.
Thus we can describe the system of the $p_z$ fermions in the blue
(thick-red) lattice as pseudo-spin up (down). By introducing
operators $\hat{c}_{\boldsymbol{i}s}$, where
$s\in\{\uparrow,\downarrow\}$ the effective Hamiltonian can be
written as $H_{\mathtt{eff}} =
T_{\mathtt{eff}}^{\parallel}\sum_s\sum_{\{\boldsymbol{i}\boldsymbol{j}\}}\hat{c}_{\boldsymbol{i}s}^\dagger\hat{c}_{\boldsymbol{j}s}
+ H_{\mathtt{int}} $ with
\begin{equation} \label{HamEff}
H_{\mathtt{int}} = V_{\uparrow\uparrow}\sum_s\sum_{\{\boldsymbol{i}\boldsymbol{j}\}} \hat{n}_{\boldsymbol{i}s}\hat{n}_{\boldsymbol{j}s}
+ V_{\uparrow\downarrow}\sum_{[\boldsymbol{i}\boldsymbol{j}]}\hat{n}_{\boldsymbol{i}\uparrow}\,\hat{n}_{\boldsymbol{j}\downarrow},
\end{equation}
where
$\hat{n}_{\boldsymbol{i}s}=\hat{c}_{\boldsymbol{i}s}^\dagger\hat{c}_{\boldsymbol{i}s}$.
For convenience we introduce
$V_{\uparrow\uparrow}=V_{p_z,p_z}(2\boldsymbol{e}_x)$ and
$V_{\uparrow\downarrow}=V_{p_z,p_z}(\boldsymbol{e}_x+\boldsymbol{e}_y)$.
Note, that now $\{.\}$ is understood as a nearest-neighbor in a
given sub-lattice. Nearest-neighbors between different sub-lattices
is denoted by $[.]$. The modified lattice constant of the
sub-lattices is $\widetilde{a}=2a$. In this way we are able to
study the system properties with the weak-coupling theory. We
investigate the emergence of triplet superconductivity between the
same pseudo-spin fermions, arising via KL mechanism \cite{Kohn} (magnetic instabilities are discussed in appendix D.). We look for Cooper pairs with chiral $p$-wave symmetry. The effective
interaction between fermions in KL mechanism in terms of the
scattering momentum $\mathbf{k-k'}=\mathbf{q}$ can be written as
\begin{align}\label{effV}
 V_{\mathtt{eff}\,\,s,s}(\boldsymbol{q})&= V_{\uparrow\uparrow} \eta_{\boldsymbol{q}}-\sum_{\boldsymbol{p}}
\left[ \left( V_{\uparrow\uparrow}^2\,\eta_{\boldsymbol{q}}^2 + V_{\uparrow\downarrow}^2\,\beta_{\boldsymbol{q}}^2\right) Q_{\boldsymbol{q},\boldsymbol{p}}  \right.  \\
&\left. -2 V_{\uparrow\uparrow}^2 \eta_{\boldsymbol{q}}\eta_{\boldsymbol{k}-\boldsymbol{p}} Q_{\boldsymbol{q},\boldsymbol{p}}
- V_{\uparrow\uparrow}^2 \eta_{\boldsymbol{k'}-\boldsymbol{p}} \eta_{\boldsymbol{k}-\boldsymbol{p}} Q_{\boldsymbol{k}+\boldsymbol{k}', \boldsymbol{p}} \right], \nonumber
\end{align}
where $Q_{\boldsymbol{q},\boldsymbol{p}} =
\frac{f(\epsilon_{\boldsymbol{p}})-f(\epsilon_{\boldsymbol{p}-\boldsymbol{q}})}{\epsilon_{\boldsymbol{p}-\boldsymbol{q}}-\epsilon_{\boldsymbol{p}}}$,
$f(\epsilon)$ is the Fermi distribution function,
$\epsilon_{\boldsymbol{p}}=2T_{\mathtt{eff}}^{\parallel}(\cos(q_x\widetilde{a})+\cos(q_y\widetilde{a}))$
is the dispersion and
$\eta_{\boldsymbol{q}}=2(\cos(q_x\widetilde{a})+\cos(q_y\widetilde{a}))$
and
$\beta_{\boldsymbol{q}}=4(\cos(q_x\widetilde{a}/2)\cos(q_y\widetilde{a}/2))$.
The summation in \eqref{effV} comes from taking into account the
second-order terms represented by diagrams shown in
Fig.~\ref{Diagrams}a. The two terms inside the first bracket in
\eqref{effV} comes from the top-left diagram in Fig~\ref{Diagrams}a,
while the next two terms comes from the top-right and bottom-left
diagrams representing vertex corrections. The last term in
\eqref{effV} comes from the bottom-right diagram in
Fig~\ref{Diagrams}a denoting exchange interactions. By performing
the integration over the momentum in the limit of $T\rightarrow 0$,
we finally get antisymmetric part of effective coupling
$\left\{V_{\mathtt{eff}}(\boldsymbol{q})\right\}_-=-\lambda(T,\mu)
(\sin(k_x\widetilde{a})\sin(k'_x\widetilde{a})+\sin(k_y\widetilde{a})\sin(k'_y\widetilde{a}))$
where
$\lambda(T,\mu)=2V_{\uparrow\uparrow}+\frac{V_{\uparrow\uparrow}^2}{\pi
T_{\mathtt{eff}}^{\parallel}}F_1(T,\mu) -
\frac{V_{\uparrow\downarrow}^2}{\pi
T_{\mathtt{eff}}^{\parallel}}F_2(T,\mu)$. Functions $F_1$ and $F_2$
originate in the second-order corrections and their detailed forms
are given in the appendix E. The point is that, due to
the Van-Hove singularity in density of states, function $F_2$
contains a logarithmic divergence. At the same time function $F_1$
is analytical due to the dressing of the density of states. This
means that there always exists finite critical $\mu$ above which the
interaction is attractive and superfluidity appears. From the BCS
theory one can get an estimate of the transition temperature $T_c$
(derivation is shown in appendix E.). In Fig.
\ref{Diagrams}b we plot the transition temperature $T_c$ as a
function of deviation $\delta n$ for example parameters discussed
previously. For $\delta n\sim 0.22$ we get $T_c\sim 0.2J_s$ ($\sim
1\,\mathrm{nK}$). Thus the ground state has a checkerboard density
pattern due to the $s$ fermions and $p$-wave superfluid $p_z$
fermions at temperature below $T_c$.

\section{Conclusions}

In conclusion, we have derived a generalized Hubbard model for
dipolar fermions in an optical lattice by taking into account higher
orbitals. We have shown that the effect of these higher orbitals
leads to new phenomena. Due to the strong interaction-dependent
hopping terms in higher orbitals, these systems can be described by
effective weakly-interacting theories. For particular parameters,
near $n\gtrsim 1/4$, we found a cross-over to the one-dimensional
physics resulting in simultaneous metallic and density wave
properties. For $J_s\ll V_{ss}(\boldsymbol{e}_x)$, the $s$ fermion
checkerboard order is given by the configuration in Fig (2)a. As
$J_s$ is increased there will be single-particle and dipole
excitations at different regions of the $n=1/4$ checkerboard crystal
similar to the one considered in Ref.~\cite{Wig} for Wigner-Hubbard
crystals due to Coloumb interaction. These excitations can induce
inter-chain tunneling at different regions and the resulting model
will be subject of future study. For other set of parameters,
$n\gtrsim 1/2$, the system can be described by a weakly interacting
Hubbard model with pseudo-spin originating from the lattice
geometry. Using the KL theory, we found a transition to the chiral
$p$-wave superfluidity due to the $p_z$ fermions without destroying
the checkerboard order created by the $s$ fermions. The parameters
used here are currently experimentally achievable.

\section{Acknowledgements}

This paper was supported by the EU STREP NAME-QUAM, IP AQUTE, ERC Grant QUAGATUA, Spanish MICINN (FIS2008-00784 and Consolider QOIT), AAII-Hubbard, and the National Science Center grant No. DEC-2011/01/D/ST2/02019. T.S. acknowledges  hospitality  from  ICFO.

\vspace{.5cm}
%\onecolumngrid
\appendix

\section{Derivation of the parameters $U_{s,p_z}$, $T_{p_z,p_z}^s$ and
$T_{p_z,p_{xz}}^s$ }

In this section we represent the on-site interaction term $U_{s,p_z}$ and interaction-induced hopping terms $T_{p_z,p_z}^s$ and
$T_{p_z,p_{xz}}^s$ in terms of the single-particle wave-function ${\cal W}_{\boldsymbol{i}\sigma}(\boldsymbol{r})$
in orbital $\sigma$ localized on site $\boldsymbol{i}$. Orbital index $\sigma=\{pml\}$ denoting $p$, $m$ and $l$
excitations in $x$, $y$, and $z$ direction respectively. Then the $s$ orbitals can be written as
${\cal W}_{\boldsymbol{i}s}(\boldsymbol{r})={w}_{{i_x}0}(x)w_{{i_y}0}(y)\phi_{0}(z)$ where ${w}_{{i_x}0}(x), {w}_{{i_y}0}(y)$ are
the lowest band one-dimensional Wannier functions and $\phi_{0}(z)$ is the ground state wave-function of the harmonic oscillator in the $z$ direction. Similarly we can
write ${\cal W}_{\boldsymbol{i}p_z}(\boldsymbol{r})={w}_{{i_x}0}(x){w}_{{i_y}0}(y)\phi_{1}(z)$ and
${\cal W}_{\boldsymbol{i}p_{xz}}(\boldsymbol{r})={w}_{{i_x}1}(x){w}_{{i_y}0}(y)\phi_{1}(z)$. Here
${w}_{{i_x}1}(x)$ is the Wannier functions in the first band and $\phi_{1}(z)$ is the first excited state of the harmonic oscillator in the $z$ direction.
For simplicity we took $\boldsymbol{j}=\boldsymbol{i}+\boldsymbol{e}_x$. From this we can write various parameters as,
\begin{widetext}
\begin{subequations}\label{U1}
\begin{align}
U_{s,p_z}&=\int \{w_{{i_x}0}(x)w_{{i_y}0}(y)\}^2 \{w_{{i_x}0}(x')w_{{i_y}0}(y')\}^2 \Phi_{1,0}(z,z') {\cal V}(\boldsymbol{r-r'}) d\boldsymbol{r}d\boldsymbol{r'} \\
T_{p_z,p_z}^s(\boldsymbol{e}_x)&=\int w_{{j_x}0}(x)w_{{i_x}0}(x)w_{{i_x}0}^2(x') \{w_{{i_y}0}(y)w_{{i_y}0}(y')\}^2\Phi_{1,0}(z,z') {\cal V}(\boldsymbol{r-r'}) d\boldsymbol{r}d\boldsymbol{r'} \\
T_{p_z,p_{xz}}^s(\boldsymbol{e}_x)&= \int w_{{j_x}1}(x)w_{{i_x}0}(x)w_{{i_x}0}^2(x') \{w_{{i_y}0}(y)w_{{i_y}0}(y')\}^2\Phi_{1,0}(z,z') {\cal V}(\boldsymbol{r-r'}) d\boldsymbol{r}d\boldsymbol{r'} \\
\Phi_{1,0}(z,z')&=\{\phi_1(z)\}^2\{\phi_0(z')\}^2 - \phi_1(z)\phi_0(z)\phi_1(z')\phi_0(z').
\end{align}
\end{subequations}
The integrations over $z,z'$ can be done analytically using convolution theorem in the momentum space. Consequently in the momentum space we get
\begin{equation}\label{U2}
V(\boldsymbol{k}_\perp)={\cal F}\left\{\int \Phi_{1,0}(z,z') {\cal V}(\boldsymbol{r-r'}) dzdz' \right\}
=\frac{2\sqrt{2\pi}D}{l_z} \left [ (k l_z)^2 - \sqrt{\frac{\pi}{2}} k l_z(1+(k l_z)^2) {\rm erfcx}\left(\frac{k l_z}{\sqrt{2}}\right) \right ],
\end{equation}
\end{widetext}
where ${\cal F}\left\{\,.\,\right\}$ denotes Fourier transform, $k=|\boldsymbol{k}_\perp|=\sqrt{k^2_x+k^2_y}$,  $l_z=(\hbar/m\Omega)^{1/2}$ is a natural harmonic oscillator length unit, and ${\rm erfcx}(x)=\exp(x^2){\rm erfc}(x)$ where ${\rm erfc}(.)$ denotes complementary error function. It is important to note that $V(\boldsymbol{k}_\perp)$ is always negative for any $\boldsymbol{k}_\perp$. This explains the appearance of the attractive on-site interaction
for any value of the confinement along the $z$ direction.
\section{Luttinger liquid description for $n>1/4$}
\begin{figure}
\includegraphics[scale=0.5]{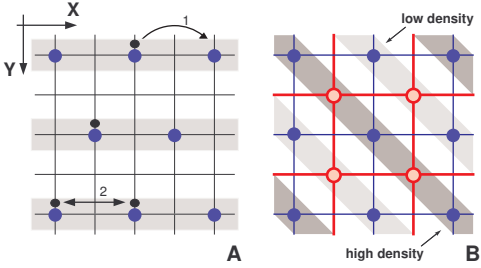}
\caption{\label{smectic} (A) Pictorial diagram for $n>1/4$. The blue spheres denote $s$-orbital fermions
and the smaller black spheres denote $p_z$ orbital fermions. The $s$-orbital fermions constitute the underlying $1/4$ checkerboard structure. The $p_z$ fermions move with the effective tunneling $T_{\mathtt{eff}}^{\parallel}$ (arrow 1) only along the shaded regions. Interaction between neighboring $p$-band fermions is equal to $V_{p_z,p_z}(2\boldsymbol{e}_x)$
(arrow 2). (B) Density-wave structure at filling $\delta n=1/4$. The dark and light shadings denote higher and lower density of the $p_z$ fermions respectively.
}
\end{figure}
As we explained in the main text, for filling $n>1/4$ and parameters $V_0=8E_R$, $D=10$, and $\hbar\Omega
\sim 14E_R$, the ground state structure is given by $1/4$ checkerboard structure formed by $s$-fermions. The $p_z$
fermions (with density $4n-1$) move in the occupied sites along $\mathbf{X}$ direction (Fig.~\ref{smectic}A).
The resulting system can be thought as  stacks of one-dimensional chains or stripes placed along $\mathbf{Y}$ without inter-chain tunnelings. The effective Hamiltonian governing the $p_z$ fermions can be
written as (see the main text):
\begin{subequations}
\begin{align}
H_{\mathtt{1D}} &=T_{\mathtt{eff}}^{\parallel} \sum_l
\sum_{\langle ij\rangle} \hat{c}^\dagger_{l,i}\hat{c}_{l,j}+H_{\mathtt{intra}}+H_{\mathtt{inter}}, \\
H_{\mathtt{intra}}&=\sum_l\sum_{i,j} V_{\mathtt{intra}}(i,j)
\hat{c}^\dagger_{l,i}\hat{c}_{l,i}\hat{c}^\dagger_{l,j}\hat{c}_{l,j}, \\
H_{\mathtt{inter}}&= \sum_{l,l'} \sum_{i,i'} V_{ll'}(i,i')\hat{c}^\dagger_{l,i}\hat{c}_{l,i}\hat{c}^\dagger_{l',i'}\hat{c}_{l',i'}.
\end{align}
\end{subequations}
The bosonized form of the intra-chain Lagrangian reads
\begin{subequations} \label{Lag0}
\begin{equation}
{\cal L}_{\rm intra}= u \int^{\pi}_{-\pi} \frac{dq_y}{2\pi} \frac{K_0}{2}
\left[\left(\frac{\partial\theta_{q_y}(x)}{\partial
t}\right)^2-\left(\frac{\partial\theta_{q_y}(x)}{\partial
x}\right)^2\right],
\end{equation}
where the Luttinger liquid parameter
\begin{equation}
K_0 =\left[\frac{2\pi T_{\mathtt{eff}}^{\parallel}\sin\tilde{k}_1+[V_{p_z,p_z}(2\boldsymbol{e}_x)+\ldots](2-\cos2\tilde{k}_{1})}{2\pi T_{\mathtt{eff}}^{\parallel}\sin\tilde{k}_1+[V_{p_z,p_z}(2\boldsymbol{e}_x)+\ldots]\cos2\tilde{k}_{1}}\right]^{1/2},
\end{equation}
and the sound velocity
\begin{eqnarray}
u^2&=&\left(2\pi T_{\mathtt{eff}}^{\parallel}\sin\tilde{k}_1+[V_{p_z,p_z}(2\boldsymbol{e}_x)+\ldots]\right)^2\nonumber\\
&-&[V_{p_z,p_z}(2\boldsymbol{e}_x)+\ldots]^2(1-\cos2\tilde{k}_{1})^2.
\end{eqnarray}
\end{subequations}
Next we include the bosonized form of the inter-chain Hamiltonian which results in the the total Lagrangian ${\cal L}_{\mathtt{1D}}={\cal L}+{\cal L}_{\rm CDW}$ where
\begin{equation} \label{Lag2}
{\cal L}= u \int^{\pi}_{-\pi} \frac{dq_y}{2\pi} \frac{K(q_y)}{2}
\left[\frac{1}{v(q_y)} \left(\frac{\partial\theta_{q_y}}{\partial t}\right)^2-v(q_y)\left(\frac{\partial\theta_{q_y}}{\partial x}\right)^2\right].
\end{equation}
Here the modified Luttinger parameter is given by
\begin{widetext}
\begin{equation}
\frac{K(q_y)}{K_0}= \left[1+4\frac{[V_{p_z,p_z}(\boldsymbol{e}_x+2\boldsymbol{e}_y)+V_{p_z,p_z}(3\boldsymbol{e}_x+2\boldsymbol{e}_y)+\ldots]\cos(q_y)+[V_{p_z,p_z}(4\boldsymbol{e}_y)+\ldots]\cos(2q_y)}{2\pi T_{\mathtt{eff}}^{\parallel}\sin\tilde{k}_f+[V_{p_z,p_z}(2\boldsymbol{e}_x)+\ldots](2-\cos(2\tilde{k}_{f1}))}\right]^{1/2},
\end{equation}
and sound velocity $v(q_y)=K(q_y)/K_0$. Inter-chain interactions induce additional charge-density wave (CDW)
perturbation, ${\cal L}_{\rm CDW}={\cal L}_{\rm CDW,1}+{\cal L}_{\rm
CDW,2}+\ldots$ with
\begin{subequations}
\begin{eqnarray}
{\cal L}_{\rm CDW,1}&=&\frac{1}{u}\sum_N
V_{p_z,p_z}((2N+1)\boldsymbol{e}_x+2\boldsymbol{e}_y)\cos[(2N+1)\tilde{k}_{1}]\sum_{l}
\cos[2\sqrt{\pi}(\theta_l-\theta_{l+1})],\\
{\cal L}_{\rm CDW,2}&=& \frac{1}{u}\sum_N
V_{p_z,p_z}(2N\boldsymbol{e}_x+4\boldsymbol{e}_y) \cos[(2N)\tilde{k}_{1}] \sum_{l}
\cos[2\sqrt{\pi}(\theta_l-\theta_{l+2})].
\end{eqnarray}
\end{subequations}
\end{widetext}
At half-filling, i.e. $\tilde{k}_1=\pi/2$, we see that ${\cal L}_{\rm CDW,1}=0$. It means that the charge-density wave instability is induced by the next-nearest neighbour inter-chain interaction. We checked that this interaction is much weaker than the tunneling $T_{\mathtt{eff}}^{\parallel}$. It means that the smectic-metal phase discussed in the paper will be stable till low enough temperature.

\section{Ground state structure of $s$-orbital fermions at $n=1/2$}
To look into the ground state of $s$-orbital fermions at $n=1/2$, we
express the average density $\langle
\hat{n}_{\boldsymbol{i},s}\rangle =(1+(-1)^{i_x+i_y}\delta)/2$,
where $\delta$ is the order parameter. We also define the
single-particle imaginary time Green functions ${\cal
G}(\boldsymbol{i}-\boldsymbol{j},\tau)=\langle {\cal T}
\hat{a}_{\boldsymbol{i},s}(\tau)\hat{a}_{\boldsymbol{j},s}^\dagger(0)\rangle$,
where ${\cal T}$ denotes time-ordering. By following the procedure
described in \cite{Kagan1, Khom} we find the following equations for
${\cal G}$ in the momentum space
\begin{subequations} \label{G1G2}
\begin{align}
\left[\omega+\mu -2V(1-\delta')\right]G_1(\boldsymbol{k},\omega)-\epsilon_{\boldsymbol{k}}G_2(\boldsymbol{k},\omega) &=1, \\
\left[\omega+\mu
-2V(1+\delta')\right]G_2(\boldsymbol{k},\omega)-\epsilon_{\boldsymbol{k}}G_1(\boldsymbol{k},\omega)
&=0,
\end{align}
\end{subequations}
where the kinetic energy $\epsilon_{\boldsymbol{k}}=2J_s\left(\cos k_xa+\cos k_ya\right)$, the
effective potential $V=\sum_{\boldsymbol{i}\neq 0} V_{ss}(\boldsymbol{i})$, and $\delta'=\delta
(\sum_{\boldsymbol{i}\in\mathtt{odd}} V_{ss}(\boldsymbol{i})-\sum_{0\neq\boldsymbol{i}\in\mathtt{even}}V_{ss}(\boldsymbol{i}))/V
$. In the position space $G_1(\boldsymbol{i})$
($G_2(\boldsymbol{i})$) is equal to ${\cal G}(\boldsymbol{i})$ for
$i_x+i_y$ even (odd). These mean field equations for $G_1$ and $G_2$ are similar to the
ones found for extended Hubbard model, with a renormalized nearest
neighbour interaction and density imbalance \cite{Kagan1, Khom}.  Then, by solving equations \eqref{G1G2}, we
find that in the strong coupling limit $\delta = 1 -
\frac{3J_s^2}{2V_{ss}^2(\boldsymbol{e}_x)}$ ($\delta \sim .98$ for $D=8$ and $\hbar\Omega=10E_R$).
Thus our assumption of a checkerboard lattice with alternative sites occupied (like the one in Fig. 2(c) in the main text) is justified.

\section{Transition temperature for Stoner Ferromagnetism and Charge-density wave instability of the $p_z$ fermions for $n>1/2$}

In this section we discuss the appearance of Stoner Ferromagnetism
and charge-density wave (CDW) instability of the $p_z$ fermions. To
do this we transform to momentum space and introduce charge
fluctuations $\rho_{\boldsymbol{q}}=\sum_{\boldsymbol{k},s}
c_{\boldsymbol{k}+\boldsymbol{q},s}^\dagger c_{\boldsymbol{k},s}$
and spin fluctuations $S_{\boldsymbol{q}}=\sum_{\boldsymbol{k},s} s
\,c_{\boldsymbol{k}+\boldsymbol{q},s}^\dagger c_{\boldsymbol{k},s}$
operators ($s\in\left\{\uparrow ,\downarrow\right\}$). Subsequently we rewrite Eq.~(4) from the main text in the
momentum space as
\begin{align}
H_{\mathtt{int}}&=\frac{1}{4} \sum_{\boldsymbol{q}} \left(2V_{\uparrow\uparrow}\eta_{\boldsymbol{q}}+V_{\uparrow\downarrow}\beta_{\boldsymbol{q}}\right) \rho_{\boldsymbol{q}}\rho_{-\boldsymbol{q}} \nonumber \\
&+\frac{1}{4} \sum_{\boldsymbol{q}}
\left(2V_{\uparrow\uparrow}\eta_{\boldsymbol{q}}-V_{\uparrow\downarrow}\beta_{\boldsymbol{q}}\right)S_{\boldsymbol{q}}S_{-\boldsymbol{q}},
\label{HINT}
\end{align}
where
$\eta_{\boldsymbol{q}}=2(\cos(q_x\widetilde{a})+\cos(q_y\widetilde{a}))$
and
$\beta_{\boldsymbol{q}}=4(\cos(q_x\widetilde{a}/2)\cos(q_y\widetilde{a}/2))$.
The system with interactions described by \eqref{HINT} can manifest
three possible magnetic instabilities: charge-density wave (CDW),
spin-density wave (SDW), and ferromagnetic instability. At $\delta
n=1/4$ (each sub-lattice is half-filled with $p_z$-orbital fermions), as
$\beta_{\boldsymbol{q}}=0$ for nesting vector $\boldsymbol{q}
\widetilde{a}=(\pm\pi,\pm\pi)$, interaction in the spin channel
becomes repulsive and therefore SDW order is absent. In the spin channel, the onset of an Stoner Ferromagnetism is given
by the divergence of susceptibility with momentum
$\boldsymbol{q}=(0,0)$. This condition can be written as $\lambda_{\rm st}\chi(0)=1$,
where $\lambda_{\rm st}=\frac{1}{2} \left [
V_{\uparrow\downarrow}\beta_{\boldsymbol{0}}-2V_{\uparrow\uparrow}\eta_{\boldsymbol{0}}
\right ]$, and the bare susceptibility $\chi(0)=\lim_{q\rightarrow 0}
\int d\mathbf{p}\,Q_{\boldsymbol{q}, \boldsymbol{p}}$ with
$Q_{\boldsymbol{q},\boldsymbol{p}} =
\frac{f(\epsilon_{\boldsymbol{p}})-f(\epsilon_{\boldsymbol{p}-\boldsymbol{q}})}{\epsilon_{\boldsymbol{p}-\boldsymbol{q}}-
\epsilon_{\boldsymbol{p}}}$. Here $f(\epsilon)$ is the Fermi
distribution function and
$\epsilon_{\boldsymbol{p}}=2T_{\mathtt{eff}}^{\parallel}(\cos(q_x\widetilde{a})+\cos(q_y\widetilde{a}))$
is the dispersion relation. Consequently, in the limit of $T \rightarrow 0 $,
$\chi(0)=\int N(\epsilon)\,\partial_\epsilon f d\epsilon$ where the
two-dimensional density of states
$$
N(\epsilon)=K(\sqrt{1-(\epsilon+\mu_p)^2/16T_{\mathtt{eff}}^{\parallel}})/
2 \pi^2 T_{\mathtt{eff}}^{\parallel},
$$
with $K(.)$ being an elliptic
integral of first kind. Substituting the density of states we get,
$\chi(0)\approx N(T)$. As $\mu\rightarrow 0$, or density of $p_z$
fermions are near $1/4$, due to the logarithmic divergence of $K$,
the transition temperature for the Stoner Ferromagnetism is given by
$$
T_{\rm st} \approx 8 T_{\mathtt{eff}}^{\parallel}\exp
\left[-\frac{2\pi^2T_{\mathtt{eff}}^{\parallel}}{\lambda_{\rm
st}}\right].
$$
The case $V_0=8E_R$, around $D\sim 8$, $\lambda_{\rm
st}/2\pi^2T_{\mathtt{eff}}^{\parallel}\sim 0.1$, corresponds to a
very low Stoner temperature $T_{\rm st}\sim 10^{-4} T_{\mathtt{eff}}^{\parallel}$.

Next we discuss the checkerboard charge-density wave structure due to the $p_z$ fermions
at $\delta n=1/4$. Due to nesting, each fermionic component will be unstable towards
CDW. In the density channel, the onset of an CDW is indicated by the divergence of susceptibility with momentum $\boldsymbol{q\tilde{a}}=(\pi,\pi)$.
The condition for $p_z$ fermion CDW can be written as
$\lambda_{\rm CDW} \chi(\pi,\pi)=-1$, where
$$
\lambda_{\rm CDW}=\frac{1}{4} \left[2V_{\uparrow\uparrow}\eta_{(\pi,\pi)}+V_{\uparrow\downarrow}\beta_{(\pi,\pi)}\right]=-2V_{\uparrow\uparrow},
$$
and the bare susceptibility
\begin{eqnarray}
\chi(\pi,\pi)&=&\lim_{\boldsymbol{q}\rightarrow (\pi,\pi), \omega
\rightarrow T} \int d\mathbf{p}
\frac{f(\epsilon_{\boldsymbol{p}})-f(\epsilon_{\boldsymbol{p}-\boldsymbol{q}})}{\omega-\epsilon_{\boldsymbol{p}-\boldsymbol{q}}+
\epsilon_{\boldsymbol{p}}} \nonumber\\
&\approx& (\log
|8T_{\mathtt{eff}}^{\parallel}/T|)^2/ 2 \pi^2
T_{\mathtt{eff}}^{\parallel}. \nonumber\\
\end{eqnarray}
Subsequently, the transition
temperature to the CDW is given by $T_{\rm
CDW}\!\approx\!8T_{2,||}\!\exp \left(-\pi
\sqrt{T_{\mathtt{eff}}^{\parallel}/V_{\uparrow\uparrow}} \right)$.
For example, when $V_0=8E_R$,
$\hbar \Omega = 10 E_R$, and $D=8$ we find that $T_{\mathrm{CDW}}
\approx 0.35 J_s$ ($\sim 2\,\mathrm{nK}$). One should note that due
to the relative shift of sub-lattices the resulting density
modulation in this phase looks like stripes rather than the standard
checkerboard structure as shown in Fig.~\ref{smectic}B.

\section{Derivation of effective interaction in the triplet channel}

In this section we derive the $p$-wave interaction from the Kohn-Luttinger effective interaction in terms
of the scattering momentum $\boldsymbol{q}=\boldsymbol{k}-\boldsymbol{k}'$. For this purpose we rewrite Eq.~(6) from the
original paper
\begin{eqnarray}\label{effV1}
V_{\mathtt{eff}\,\,s,s}(\boldsymbol{q})&=&V_{\uparrow\uparrow} \eta_{\boldsymbol{q}}-\sum_{\boldsymbol{p}}
\left[ \left( V_{\uparrow\uparrow}^2\,\eta_{\boldsymbol{q}}^2
+ V_{\uparrow\downarrow}^2\,\beta_{\boldsymbol{q}}^2\right) Q_{\boldsymbol{q},\boldsymbol{p}} \right. \nonumber\\
  &-& \left. 2 V_{\uparrow\uparrow}^2 \eta_{\boldsymbol{q}}\eta_{\boldsymbol{k}-\boldsymbol{p}} Q_{\boldsymbol{q},\boldsymbol{p}}
- V_{\uparrow\uparrow}^2 \eta_{\boldsymbol{k'}-\boldsymbol{p}} \eta_{\boldsymbol{k}-\boldsymbol{p}} Q_{\boldsymbol{k}+\boldsymbol{k}', \boldsymbol{p}} \right], \nonumber\\
&&
\end{eqnarray}
where $Q_{\boldsymbol{p},\boldsymbol{q}}=\frac{f(\epsilon(\boldsymbol{p}))-f(\epsilon(\boldsymbol{p-q}))}{\epsilon(\boldsymbol{p-q})-\epsilon(\boldsymbol{p})}$ with $f(.)$ being the Fermi distribution function.

First we put the expression of $\eta_{\boldsymbol{q}}=2(\cos(q_x\tilde{a})+\cos(q_y\tilde{a}))$ and
$\beta_{\boldsymbol{q}}=4(\cos(q_x\tilde{a}/2)\cos(q_y\tilde{a}/2))$ back to \eqref{effV1}. As we are interested in $p$-wave interaction, after expanding \eqref{effV1} in terms of the momenta $k_x$, $k'_x$ ,$k_y$, $k'_y$, we keep terms proportional to $\sin k_x\widetilde{a}\sin k'_x\widetilde{a}+\sin k_y\widetilde{a}\sin k'_y\widetilde{a}$. In this way we get
\begin{widetext}
\begin{align} \label{EQ1}
\{V_{\mathtt{eff}}(\boldsymbol{q})\} &= \left ( 2V_{\uparrow\uparrow} - \sum_{\boldsymbol{p}}Q_{\boldsymbol{q},\boldsymbol{p}}
[4(V_{\uparrow\downarrow})^2-8(V_{\uparrow\uparrow})^2\{\cos (k_x\tilde{a}-p_x\tilde{a}) + \cos (k_y\tilde{a}-p_y\tilde{a})\}] \right) (\sin k_x\tilde{a}\sin k'_x\tilde{a}+\sin k_y\tilde{a}\sin k'_y\tilde{a}) \nonumber \\
&+ 4(V_{\uparrow\uparrow})^2
\sum_{\boldsymbol{p}}Q_{\boldsymbol{q},\boldsymbol{p}} (\sin k_x\tilde{a}\sin k'_x\tilde{a}\sin^2 p_x\tilde{a}+\sin k_y\tilde{a}\sin k'_y\tilde{a}\sin^2p_y\tilde{a})
\end{align}
\end{widetext}
where $\boldsymbol{q}=\boldsymbol{k-k'}$. By converting sums to integrals in \eqref{EQ1} we have to compute terms of the form
$\int d\boldsymbol{p}Q_{\boldsymbol{q},\boldsymbol{p}}$, $\int d\boldsymbol{p} \sin^2p_xa_b Q_{\boldsymbol{q},\boldsymbol{p}}$, and
$\int d\boldsymbol{p}\cos p_xa_b Q_{\boldsymbol{q},\boldsymbol{p}}$. In the limit of vanishing temperature $T \rightarrow 0$ we approximate all integrals $\int d\boldsymbol{p} G(\boldsymbol{p})\frac{f(\epsilon(\boldsymbol{p}))-f(\epsilon(\boldsymbol{p-q}))}{\epsilon(\boldsymbol{p-q})-\epsilon(\boldsymbol{p})}\approx \int N_{\mathtt{eff}}(\epsilon,\mu)\partial_\epsilon f d\epsilon$ for arbitrary function $G(\boldsymbol{p})$. The effective density of states reads $N_{\mathtt{eff}}(\epsilon,\mu)=\int d\boldsymbol{k}
G(\boldsymbol{k}) \delta(\epsilon-\epsilon_{\boldsymbol{k}})$. We see that the only first term inside the third bracket in \eqref{EQ1} is not dressed by $\cos(p_x\tilde{a})$ or $\sin(p_x\tilde{a})$. Hence the effective density of states for this term contains Van-Hove singularity. All other terms with in \eqref{EQ1}, due to the dressing  by
$\cos(p_x\tilde{a})$ or $\sin(p_x\tilde{a})$, are analytic. For convenience, we re-express
$\left\{V_{\mathtt{eff}}(\boldsymbol{q})\right\}_-=-\lambda(T,\mu) (\sin(k_x\widetilde{a})\sin(k'_x\widetilde{a})+\sin(k_y\widetilde{a})\sin(k'_y\widetilde{a}))$,
where $\lambda(T,\mu)=2V_{\uparrow\uparrow}+\frac{V_{\uparrow\uparrow}^2}{\pi T_{\mathtt{eff}}^{\parallel}}F_1(T,\mu) -
 \frac{V_{\uparrow\downarrow}^2}{\pi T_{\mathtt{eff}}^{\parallel}}F_2(T,\mu)$. The functions $F_1$ and $F_2$ are given by
 \begin{subequations}
\begin{align}
F_1&=\frac{4}{\pi} F\left(1-(T+|\mu|)^2/(4T_{\mathtt{eff}}^{\parallel})^2\right), \\
F_2&=\frac{2}{\pi} K\left([1-(T+|\mu|)^2/(4T_{\mathtt{eff}}^{\parallel})^2\right),
\end{align}
\end{subequations}
where $F(x)=E(x)-(1-x)K(x)$ with $K(.)$ being the Elliptic integral of
the first kind and $E(.)$ is the Elliptic integral of the second kind. Then we can write the BCS
equation for the transition temperature $T_c$ as \cite{Rob}
\begin{equation}
V_p N_{\mathtt{eff}}(0, \mu)\log \left|\left(1-\left[\frac{\mu}{4T_{\mathtt{eff}}^{\parallel}}\right ]^2 \right)^{1/2}\frac{4T_{\mathtt{eff}}^{\parallel}}{T_c} \right |=1,
\end{equation}
where the effective density of
state is given by
\begin{eqnarray}
N_{\mathtt{eff}}(\epsilon,\mu)&=&\sum_{\boldsymbol{k}}\delta(\epsilon+\mu-\epsilon_{\boldsymbol{k}})
\sin^2(k_xa_b) \nonumber\\
&\approx & F\left(1-(\epsilon+|\mu|)^2/(4T_{\mathtt{eff}}^{\parallel})^2
\right)/\pi^2T_{\mathtt{eff}}^{\parallel}.
\end{eqnarray}
As discussed in the main text, the ground state has a checkerboard density
pattern due to the $s$ fermions and $p$-wave superfluid $p_z$
fermions at temperature below $T_c$. From previous section, we see that at $\delta n=1/4$ (half filled
$p_z$ fermions for the red and blue lattices), the transition
temperatures for the $p_z$ fermion CDW and $p_z$ fermion superfluidity are
similar. This will result in a competition or co-existance between both
instabilities for the $p_z$ orbital fermions. A detailed account of such scenario is beyond this
scope of this paper.

\end{document}